\documentclass[onecolumn]{IEEEtran}

\usepackage{graphicx}
\usepackage{url}

\title{Adventures with Grace Hopper AI Super Chip and the National Research Platform}
\author{
    \IEEEauthorblockN{J. Alex Hurt\IEEEauthorrefmark{1}, Grant J. Scott\IEEEauthorrefmark{1}, Derek Weitzel\IEEEauthorrefmark{2}, Huijun Zhu\IEEEauthorrefmark{2}}
    \\
    \IEEEauthorblockA{\IEEEauthorrefmark{1} University of Missouri - Columbia}
    \\
    \IEEEauthorblockA{\IEEEauthorrefmark{2} University of Nebraska - Lincoln}
}
\begin{document}

\maketitle
\begin{abstract}
  The National Science Foundation (NSF) funded National Research Platform (NRP) is a hyper-converged cluster of nationally and globally interconnected heterogeneous computing resources.
  The dominant computing environment of the NRP is the x86\_64 instruction set architecture (ISA), often with graphics processing units (GPUs).
  Researchers across the nation leverage containers and Kubernetes to execute high-throughput computing (HTC) workloads across the heterogeneous cyberinfrastructure with minimal friction and maximum flexibility.
  As part of the NSF-funded GP-ENGINE project, we stood up the first server with an NVIDIA Grace Hopper AI Chip (GH200), an alternative ARM ISA, for the NRP.
  This presents challenges, as containers must be specifically built for ARM versus x86\_64.
  Herein, we describe the challenges encountered, as well as our resulting solutions and some relevant performance benchmarks.
  We specifically compare the GH200 to A100 for computer vision workloads, within compute nodes in the NRP.
\end{abstract}

\section{Introduction}\label{sect:intro}
The Great Plains Network (GPN) is a regional optical network (RON) formed by state research and education networks (RENs) whose members include academic institutions and state networks in South Dakota, Nebraska, Kansas, Oklahoma, Missouri, Arkansas, and Iowa.
The Great Plains region is characterized by a rural sparsity of population and a general shortage of highly trained cyberinfrastructure professionals (CIPs).
To overcome these challenges, the GPN community operates as a collaborative, knowledge-sharing community.
In 2019, the Great Plains CyberTeam \cite{GPCT} was established by NSF award.
This formalized many of the region's collaborations and mutual support patterns, which then facilitated additional projects for distributed regional computing.

\begin{figure}[h]
  \centering
  \caption{(a) GP-ENGINE sites, each with at least 4x A100 GPUs on NRP. University of Missouri (MU) hosts two 4x A100 nodes and the NRP's first Grace Hopper node. (b) Typical computer vision workflow in the NRP.}
  \begin{tabular}{cc}
    \includegraphics[width=0.40\linewidth]{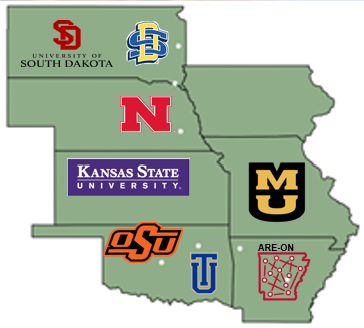} &
    \includegraphics[width=0.25\linewidth]{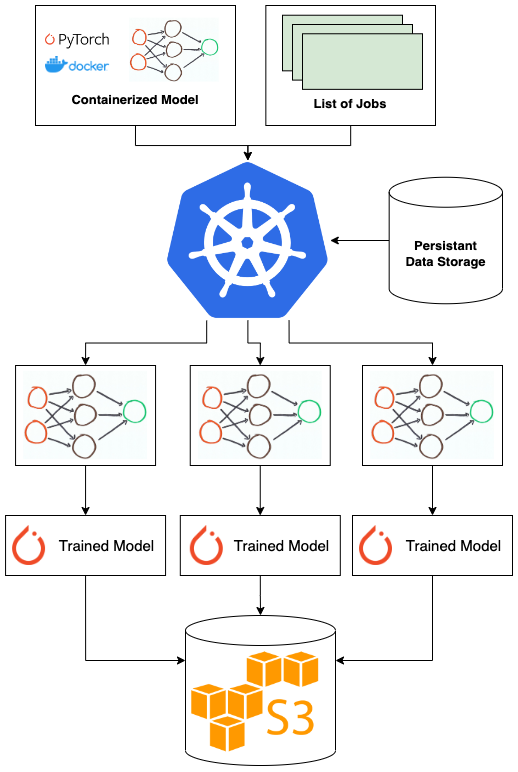}     \\
    (a)                                                                & (b)
  \end{tabular}
  \label{fig:GPENGINE}
\end{figure}

The GPN has now developed extensive experience deploying multi-state distributed compute nodes for high-throughput computing (HTC).
The Great Plains Augmented Regional Gateway to the Open Science Grid (GP-ARGO) \cite{GPArgo} established a network of HTC within the GPN by distributing computational gateways to OSG.
More recently, the Great Plains Extended Network of GPUs for Interactive Experimenters (GP-ENGINE) \cite{GPENGINE} built upon this collaborative energy, leveraging our existing GP-CyberTeam to train researchers to accelerate their codes and migrate them to HTC resources.
GP-ENGINE supports cross-institutional computational and data-intensive research around the region through the development of specific cyberinfrastructure (CI) resources and workforce training.
The project’s primary goal is to help researchers transition nascent ideas and workbench codes into GPU-accelerated code, which can become ``HTC Ready'' modules for national HTC resources such as the National Research Platform (NRP).
The systems reported in this manuscript are part of the GP-ENGINE, which is a subset of the NRP with nodes located at the GPN sites (see Fig.~\ref{fig:GPENGINE}).
The primary nodes of GP-ENGINE are servers with 4x 80GB NVIDIA A100, 320GB GPU RAM, and 27,648 CUDA cores.
Additionally, each of these nodes contains 2x AMD 7713 with 128 Cores and 256 Threads, 1 TB RAM, and 2x 25Gbe network cards.
As of writing, since October 1, 2023, GP-ENGINE has provided the following research support in the US: 101 research groups supported, 115,000 GPU hours, and 1.6 million CPU hours.
Importantly, GP-ENGINE brought online the NRP's first Grace Hopper composite chip (GH200), with the tightly coupled ARM CPU (72 cores) and H100 GPU (14,592 CUDA cores), and 600 GB total RAM.
\section{Computational Performance Comparison GH200 vs A100}\label{sect:perf}

We now move to the presentation and analysis of experimental results utilizing the GH200 for training deep neural networks (DNNs) to perform computer vision (CV) tasks.
In this short paper, we only discuss computational efficiency. %
Our evaluation of the GH200 includes an analysis of its throughput and efficiency performance on three CV tasks: image classification, object detection, and semantic segmentation.
For the experiments shown here, we utilize both standard convolutional neural networks (CNNs), as as well Transformer-based feature extraction methods.

The experimental design for each experiment will be to train models with identical hyper-parameters, such as training length, learning rate, and network optimization algorithm, on heterogeneous machines and compare the throughput and overall training and/or inference time to complete the CV task.
We compare the GH200's performance with two other hardware configurations: a single 80GB NVIDIA A100 (1xA100) and four 80GB NVIDIA A100s utilizing Distributed Data Parallelism (4xA100 DDP).

In the sections below, we describe in more detail the performance of the GH200 against the competing hardware configurations for each CV task.
\subsection{Image Classification with Deep Neural Networks}\label{sect:perf:class}

We begin our experimentation with DNNs training for image classification.
DNNs trained for image classification are typically composed of a feature extraction network (CNN or Transformer), followed by a flattening layer that transforms the 2-D features into a 1-D feature vector, which is then fed into at least one fully-connected layer, which then uses the features for classification, usually with a SoftMax layer to normalize classification scores between 0 and 1.
While feature extraction architectures range in complexity and design, for these experiments, we utilize three popular feature extraction architectures: ResNet, Vision Transformer, and SWIN Transformer.

\subsubsection{Image Classification Architectures}

\begin{table}[!t]
\centering
  \caption{Comparison of Times (in minutes) for training classifiers between 1xA100, 4xA100, and GH200 on (a) CIFAR-100 and (b) Food101.}
\begin{tabular}{cc}
  \begin{tabular}{|l|c|c|c|}
    \hline
    \textbf{Model} & \textbf{1xA100} & \textbf{4xA100} & \textbf{GH200} \\
    \hline
    ResNet50       & 36.67           & 19.74           & 24.81          \\
    ResNet152      & 63.87           & 27.08           & 38.77          \\
    ViT-L          & 46.20           & 16.39           & 26.93          \\
    ViT-H          & 1948.01         & 509.7           & 890.3          \\
    SWIN-T         & 52.76           & 15.83           & 34.5           \\
    SWIN-B         & 94.48           & 27.13           & 56.71          \\
    \hline
  \end{tabular}
& 
  \begin{tabular}{|l|c|c|c|}
    \hline
    \textbf{Model} & \textbf{1xA100} & \textbf{4xA100} & \textbf{GH200} \\
    \hline
    ResNet50       & 89.01           & 23.75           & 44.01          \\
    ResNet152      & 111.30          & 35.45           & 67.82          \\
    ViT-L          & 83.63           & 24.96           & 41.12          \\
    ViT-H          & 2947.94         & 762.03          & 1355.09        \\
    SWIN-T         & 96.66           & 25.89           & 57.39          \\
    SWIN-B         & 167.65          & 44.04           & 96.53          \\
    \hline
  \end{tabular}
\\
(a) CIFAR-100 & (b) Food101 \\
\end{tabular}
  \label{table:classification-training}
\end{table}

ResNet \cite{resnet} is a residual CNN and one of the most popular CNN feature extractors in CV research.
Its key design philosophy is the addition of residual connections between convolutional layers that enable it to perform superior feature extraction while limiting the effect of the vanishing gradient problem.
For this research, we utilize ResNet50 and ResNet152, which require the optimization of 25.6 million and 60.2 million learnable parameters, respectively.

The Vision Transformer (ViT) \cite{vit} is a transformer-backbone that utilizes a global self-attention mechanism to enable the model to learn the optimal features and encode global context into the learned features.
The SWIN architecture \cite{swin} improves upon this design by using hierarchical shifted windows of 7x7 self-attention that enable better multi-scale feature extraction while also improving computational efficiency.
For this research, we utilize the Large and Huge (306.5 million and 633.5 million learnable parameters) sizes for ViT, and the Tiny and Base (28.3 million and 87.8 million learnable parameters) sizes for the SWIN Transformer.

\subsubsection{Image Classification Datasets}
The datasets selected for Image Classification experiments include CIFAR-100 \cite{cifar100} and Food101 \cite{food101}.
CIFAR-100 is a 32x32 pixel RGB CV dataset containing 60,000 samples spanning 100 classes.
Meanwhile, Food101 contains image dimensions as large as 512 pixels in 101,000 samples belonging to 101 Food categories.
Finally, the popular ImageNet dataset \cite{imagenet} is used for inference benchmarks.

\subsubsection{Training Benchmarks}

Each of the six model configurations are applied to the two selected datasets in three hardware configurations: GH200, 1x A100, and 4x A100 (DDP), and the results are shown in Table \ref{table:classification-training}.
Results on CIFAR-100 show that GH200 is 32.3\% and 39.3\% faster than 1x A100 on ResNet50 and ResNet152, respectively, but 20.4\% and 30.2\% slower than 4x A100 on those same networks.
This trend continues for the transformer-based architectures, where 4xA100 is 42.7\% faster than GH200 on our most computationally expensive model, ViT-H, but GH200 is again 54.3\% faster than its single A100 counterpart.
These same takeaways continue with the larger dataset, Food101, as the GH200 is again more than 50\% faster than the A100 (50.6\%), but significantly behind the distributed 4x A100 configuration (46.0\%).
Overall, the training benchmarks for image classification show a consistent 35-55\% increase in throughput for the GH200 against a single A100, but a lag of 20-55\% behind four A100s, depending on the dataset and model configuration.

\begin{figure*}[!t]
\begin{center}
  \centering
  \caption{Inference Analysis (a) Time (s) and (b) Images Per Second.}
  \begin{tabular}{cc}
    \includegraphics[width=0.45\linewidth]{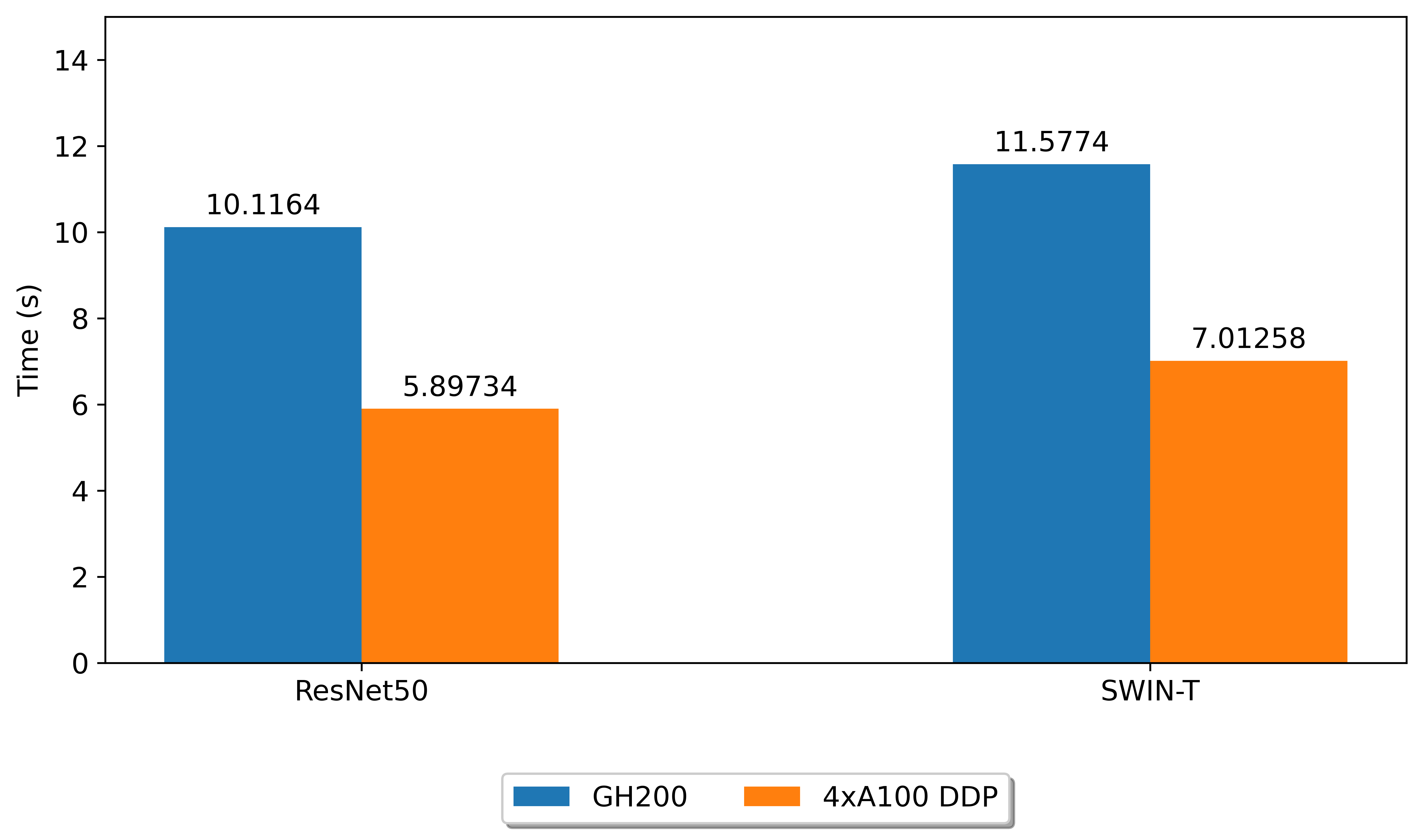} &
    \includegraphics[width=0.45\linewidth]{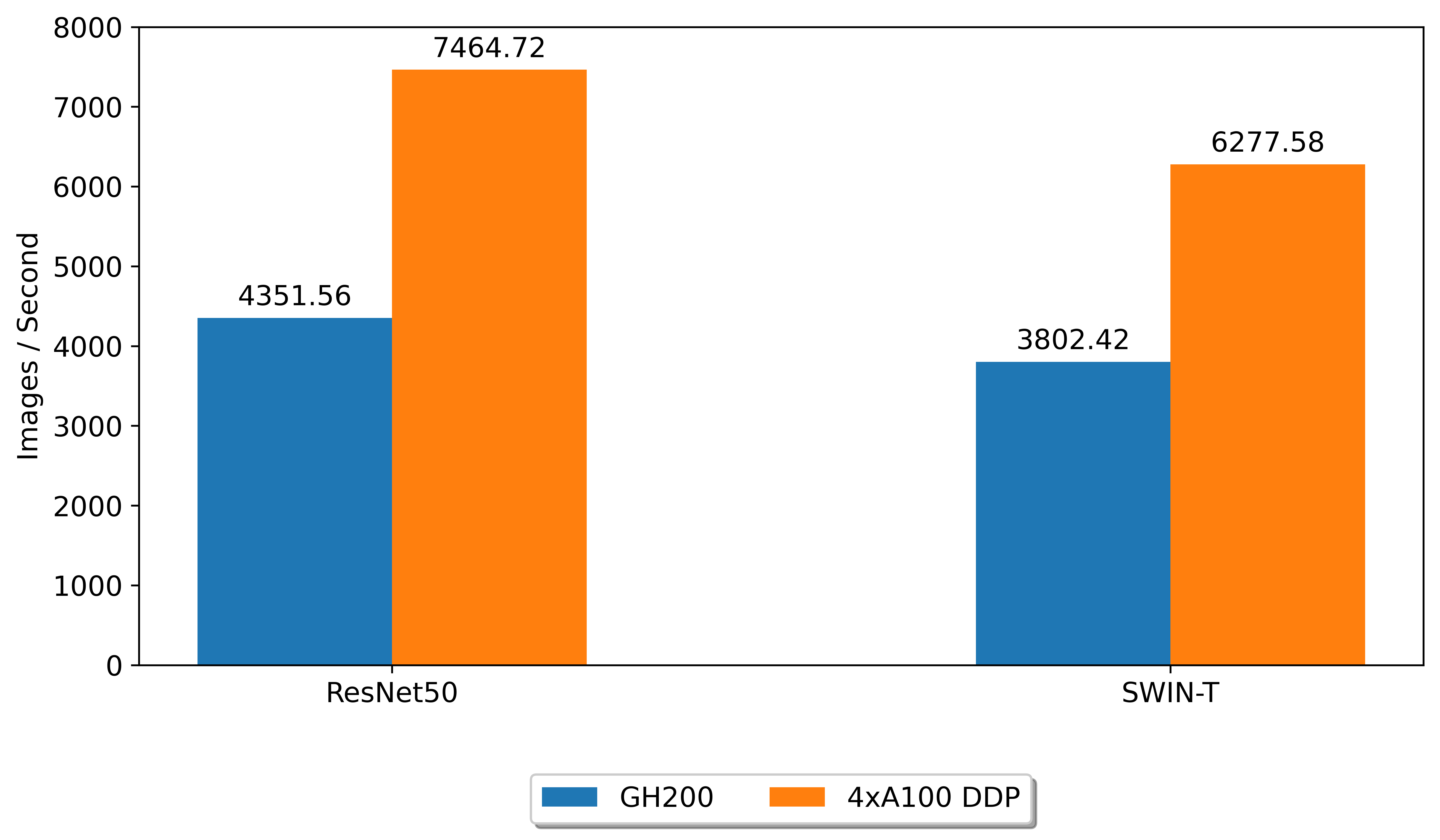}        \\
    (a) Time(s)                                                                             & (b) Images Per Second \\
  \end{tabular}
  \label{fig:classification-inference}
\end{center}
\end{figure*}

\subsubsection{Inference Benchmarks}
Our final analysis of GH200 throughput for the image classification task compares the inference performance of the 4x A100 configuration with the GH200.
We run inference using two networks by applying ImageNet pretrained models from TorchVision \cite{torchvision2016} to the ImageNet validation dataset of 50,000 images, and record the total runtime and images processed per second.
The results of these benchmarks, shown in Fig.~\ref{fig:classification-inference}, reveal a performance difference of 41.7\% with ResNet50, which is reasonably similar to the 46\% difference found in training on Food101.
However, the inference benchmark on SWIN-T shows a performance difference of only 39.4\%, much lower than the 54.9\% difference observed during training of SWIN-T on Food101, indicating that GH200 is markedly more performant for model inference on transformers than it is for CNN-based models relative to 4x A100.

\subsection{Object Detection with Deep Neural Networks}\label{sect:perf:bbox}
\begin{table}[!t]
    \centering
  \caption{Comparison of Times (in hours) for training object detection on COCO and \\ semantic segmentation models on CityScapes between 1xA100, 4xA100, and GH200}
  \begin{tabular}{|l|l|l|c|c|c|}
    \hline
    \textbf{Model} & \textbf{Backbone} & \textbf{Dataset} & \textbf{1xA100} & \textbf{4xA100} & \textbf{GH200} \\
    \hline
    DINO           & SWIN-L            & COCO             & 125.00          & 37.77           & 34.38          \\
    Faster R-CNN   & ResNet-50         & COCO             & 4.59            & 1.08            & 11.84          \\
    \hline
    SegFormer      & MIT-B4            & CityScapes       & 10.76           & 2.75            & 4.53           \\
    DeepLab V3+    & ResNet-101        & CityScapes       & 4.19            & 1.18            & 2.78           \\
    \hline
  \end{tabular}
  \label{table:detection-training}
\end{table}

Our next set of experiments evaluates the GH200's performance for the Object Detection task.
The experimental setup is identical to the previous section: We will select a set of models, train identical versions of them on three machine configurations, and compare the relative performance in time to complete the training of the models.
For these experiments, we utilize the popular COCO dataset \cite{coco} and select two models from the MMDetection model zoo: Faster R-CNN with a ResNet-50 backbone and Feature Pyramid Network, and a DINO model with a SWIN-L backbone.

\subsubsection{Object Detection Architectures}
Our selected Convolutional-based object detector is the Faster R-CNN \cite{faster_rcnn} architecture utilizing the same ResNet-50 \cite{resnet} architecture seen in the Image Classification experiments.
The detection architecture used for this model, Faster R-CNN, utilizes a region proposal network (RPN) to nominate potential regions of interest from convolutional feature maps.
The original Faster R-CNN showed poor performance in applications requiring scale robustness, so we choose to utilize a model with a Feature Pyramid Network (FPN) \cite{FPN}, which increases scale robustness throughout feature extraction.
The transformer-based model used for these experiments is the DINO \cite{dino} detector utilizing a SWIN-L backbone.
The detection methodology itself, DINO, is a transformer-based detection method that improves upon previous end-to-end transformer architectures such as DETR \cite{detr}.
The number of learnable parameters for our backbones, ResNet-50 and SWIN-L, are 25.6 million and 197 million, respectively.

\subsubsection{Training Benchmarks}

The two selected models are trained with the three selected machine configurations, and results are shown in Table~\ref{table:detection-training}, Rows 1 and 2.
While in image classification, we observed a consistent trend of 4x A100 being 20-50\% better than GH200 with 1x A100 performing 20-50\% worse than the GH200, our results here are a direct contradiction with those conclusions.
For the DINO model with SWIN-L, we see that the GH200 is not only a staggering 72.5\% faster than 1xA100, but also nearly 10\% (9.9\%) faster than 4xA100 in training times.
This performance trend significantly changes when our detection architecture changes from transformer-based DINO to convolutional-based Faster R-CNN, as GH200 drops from the fastest performer to the slowest.
In detection with Faster R-CNN, GH200 is outperformed by 1x A100 by 58\% and by 4x A100 by over 90\% (90.9\%).
These results show the sensitivity of hardware configurations to the computational complexity of the model, as GH200 performs excellently using modern architectures, but is outperformed by the previous generation of GPU when using older architectures.

\subsection{Semantic Segmentation with Deep Neural Networks}\label{sect:perf:seg}

Our final set of experiments for the GH200 evaluates its performance and throughput for training DNNs for the semantic segmentation task. 
As with previous experiments, we will train two model architectures, one convolutional and one transformer-based, on three distinct hardware configurations: GH200, 1x A00, and 4x A100.
Our selected dataset for these experiments is the popular CityScapes dataset \cite{cityscapes}, and our selected models and pre-trained weights are taken from the MMSegmentation framework: SegFormer with MIT-B4 backbone, and DeepLabV3+ with ResNet-101 backbone.

\subsubsection{Segmentation Architectures}

DeepLabV3+ \cite{deeplabv3plus} is a convolutional-based segmentation architecture that improves upon the previous DeepLabV3 \cite{deeplabv3} architecture by adding a decoder module along object boundaries, as well as utilizing depth-separable convolutions.
The backbone selected for this architecture is the 101-layer variant of ResNet, ResNet-101 	
(44.5 million parameters).
SegFormer \cite{segformer}, our transformer-based segmentation algorithm, is an efficient transformer architecture utilizing both global and local attention with hierarchical structure while forgoing the positional encoding found in ViT and SWIN to improve computational throughput. 
SegFormer defines its own backbone, and we choose the B4 size (64.1 million parameters) for these experiments.

\subsubsection{Training Benchmarks}
The results from segmentation training benchmarks are shown alongside the previously discussed detection benchmarks in Table~\ref{table:detection-training}.
Segmentation experimentation is reminiscent of classification, in that we see a consistent trend of GH200 outperforming 1xA100 while falling short of the performance of 4xA100.
For SegFormer and DeepLabV3+, the margin of GH200's lead over 1xA100 is 57.9\% and 33.7\%, respectively.
In turn, GH200 is then outperformed by 4xA100 by margins of 39.3\% and 57.6\% on these same networks, reinforcing the results from classification.

\section{Conclusion and Future Work}\label{sect:conclusion}

This short paper detailed the performance of the Grace Hopper (GH200) compared to 1x and 4x A100 compute configurations on the NRP for select computer vision tasks using publicly available benchmark datasets.
These tasks include classification, object detection, and semantic segmentation, each evaluating convolutional and transformer-based deep neural networks.
\section*{Acknowledgments}
  Computing resources for this research have been supported by the NSF National Research Platform and NSF OAC Award \#2322218 (GP-ENGINE).

\bibliographystyle{IEEEbib}
\bibliography{gp-engine.bib}

\begin{thebibliography}{10}

\bibitem{GPCT}
Carrie Brown, Ryan Johnson, Kate Adams, Kevin Brandt, Adison Kleinsasser, James Deaton, and Timothy Middelkoop,
\newblock ``Great plains cyberteam: A regional mentor approach to cyberinfrastructure workforce development and advancement,''
\newblock in {\em Practice and Experience in Advanced Research Computing}, pp. 456--460. 2020.

\bibitem{GPArgo}
Daniel Andresen, Timothy Middelkoop, Pratul Agarwal, Stephen Wheat, and Ryan Johnson,
\newblock ``Gp-argo: The great plains augmented regional gateway to the open science grid,'' 2020.

\bibitem{GPENGINE}
J.~Alex Hurt, Derek Weitzel, Daniel Andresen, Brian Burkhart, and Paul Kern,
\newblock ``Great plains extended network of gpus for interactive experimenters,'' 2023.

\bibitem{resnet}
Kaiming He, Xiangyu Zhang, Shaoqing Ren, and Jian Sun,
\newblock ``Deep residual learning for image recognition,''
\newblock in {\em Proceedings of the IEEE conference on computer vision and pattern recognition}, 2016, pp. 770--778.

\bibitem{vit}
Alexey Dosovitskiy, Lucas Beyer, Alexander Kolesnikov, Dirk Weissenborn, Xiaohua Zhai, Thomas Unterthiner, Mostafa Dehghani, Matthias Minderer, Georg Heigold, Sylvain Gelly, et~al.,
\newblock ``An image is worth 16x16 words: Transformers for image recognition at scale,''
\newblock {\em arXiv preprint arXiv:2010.11929}, 2020.

\bibitem{swin}
Ze~Liu, Yutong Lin, Yue Cao, Han Hu, Yixuan Wei, Zheng Zhang, Stephen Lin, and Baining Guo,
\newblock ``Swin transformer: Hierarchical vision transformer using shifted windows,''
\newblock in {\em Proceedings of the IEEE/CVF International Conference on Computer Vision}, 2021, pp. 10012--10022.

\bibitem{cifar100}
Alex Krizhevsky, Geoffrey Hinton, et~al.,
\newblock ``Learning multiple layers of features from tiny images,''
\newblock 2009.

\bibitem{food101}
Lukas Bossard, Matthieu Guillaumin, and Luc Van~Gool,
\newblock ``Food-101 -- mining discriminative components with random forests,''
\newblock in {\em European Conference on Computer Vision}, 2014.

\bibitem{imagenet}
Jia Deng, Wei Dong, Richard Socher, Li-Jia Li, Kai Li, and Li~Fei-Fei,
\newblock ``Imagenet: A large-scale hierarchical image database,''
\newblock in {\em 2009 IEEE conference on computer vision and pattern recognition}. Ieee, 2009, pp. 248--255.

\bibitem{torchvision2016}
TorchVision maintainers and contributors,
\newblock ``Torchvision: Pytorch's computer vision library,'' \url{https://github.com/pytorch/vision}, 2016.

\bibitem{coco}
Tsung-Yi Lin, Michael Maire, Serge Belongie, James Hays, Pietro Perona, Deva Ramanan, Piotr Doll{\'a}r, and C~Lawrence Zitnick,
\newblock ``Microsoft coco: Common objects in context,''
\newblock in {\em European conference on computer vision}. Springer, 2014, pp. 740--755.

\bibitem{faster_rcnn}
Shaoqing Ren, Kaiming He, Ross Girshick, and Jian Sun,
\newblock ``Faster r-cnn: towards real-time object detection with region proposal networks,''
\newblock {\em IEEE transactions on pattern analysis and machine intelligence}, vol. 39, no. 6, pp. 1137--1149, 2016.

\bibitem{FPN}
Tsung-Yi Lin, Piotr Doll{\'a}r, Ross Girshick, Kaiming He, Bharath Hariharan, and Serge Belongie,
\newblock ``Feature pyramid networks for object detection,''
\newblock in {\em Proceedings of the IEEE conference on computer vision and pattern recognition}, 2017, pp. 2117--2125.

\bibitem{dino}
Hao Zhang, Feng Li, Shilong Liu, Lei Zhang, Hang Su, Jun Zhu, Lionel~M Ni, and Heung-Yeung Shum,
\newblock ``Dino: Detr with improved denoising anchor boxes for end-to-end object detection,''
\newblock {\em arXiv preprint arXiv:2203.03605}, 2022.

\bibitem{detr}
Nicolas Carion, Francisco Massa, Gabriel Synnaeve, Nicolas Usunier, Alexander Kirillov, and Sergey Zagoruyko,
\newblock ``End-to-end object detection with transformers,''
\newblock in {\em European conference on computer vision}. Springer, 2020, pp. 213--229.

\bibitem{cityscapes}
Marius Cordts, Mohamed Omran, Sebastian Ramos, Timo Rehfeld, Markus Enzweiler, Rodrigo Benenson, Uwe Franke, Stefan Roth, and Bernt Schiele,
\newblock ``The cityscapes dataset for semantic urban scene understanding,''
\newblock in {\em Proceedings of the IEEE conference on computer vision and pattern recognition}, 2016, pp. 3213--3223.

\bibitem{deeplabv3plus}
Liang-Chieh Chen, Yukun Zhu, George Papandreou, Florian Schroff, and Hartwig Adam,
\newblock ``Encoder-decoder with atrous separable convolution for semantic image segmentation,''
\newblock in {\em Proceedings of the European conference on computer vision (ECCV)}, 2018, pp. 801--818.

\bibitem{deeplabv3}
Liang-Chieh Chen, George Papandreou, Florian Schroff, and Hartwig Adam,
\newblock ``Rethinking atrous convolution for semantic image segmentation,''
\newblock {\em arXiv preprint arXiv:1706.05587}, 2017.

\bibitem{segformer}
Enze Xie, Wenhai Wang, Zhiding Yu, Anima Anandkumar, Jose~M Alvarez, and Ping Luo,
\newblock ``Segformer: Simple and efficient design for semantic segmentation with transformers,''
\newblock {\em Advances in neural information processing systems}, vol. 34, pp. 12077--12090, 2021.

\end{thebibliography}

\end{document}